\documentstyle[dina4p,11pt,epsf]{article}
\begin{document}
\begin{center}
\Large{Cosmological Solution for the Problem of Rotation Velocities in
Galaxies}
\end{center}
\begin{center}
Jose Luis Rosales; Departamento de Fisica Teorica, U.A.M. Cantoblanco 28049
Madrid
\end{center}
\begin{quote}

A solution for the problem of understanding observed rotation curves
in galaxies without the introduction of dark matter halos is presented.
This solution has been obtained upon considering the
distribution of masses in the expanding universe,  then, having a cosmological
character. A formal limiting radius for galaxies depending on cosmological
parameters
is given. The empirical conclusions derived from the theory of M. Milgrom
and J. Bekenstein arise as direct consequences of the present
approach without any need of drastically modifying newtonian dynamics.
\end{quote}

PACS numbers:98.80.Hw, 95.35.+d, 98.62.Dm

\vspace{2 mm}
Since the discovery of  motions of stars in the galaxies and their
gravitational interactions derived from them, it seems evident
that most part of the matter
in the universe should be dark. This comes from the measurements of
orbital velocities of stars as a function of the distance from the centre of
the
galaxy. There are no important regions where the velocities fall off with
distance from the centre, as would be expected if the mass were practically
concentrated. Moreover, also from
rotation curves, spiral galaxies seem  to have no limits;
in fact, their deduced masses increase endlessly with
distance\cite{kn:Ostriker}.
Now, the question is whether this behaviour (of not having a particular limit
for every particular galaxy) might (or not) be of cosmological nature.

In this letter a different point of view will be introduced. We shall
contemplate the possibility
that the observed gravitational behaviour
of a particle of mass $M$ in the expanding universe be different from the
one expected if we were using standard static newtonian theory;
therefore, in order to understand such a behaviour
we must take into account, additionaly, cosmological boundary conditions.
This is, somehow, a machian view of gravity, i.e., there should be an
important relationship between the behaviour and properties of the universe as
a whole and the corresponding dynamics of small bodies in relative motion
within it \cite{kn:Sciama}.

Let us, then, obtain a cosmological correction to the newtonian theory of
gravity
which would explain the disagreement between observed  and predicted motions;
moreover, this correction should be of universal application and, then,
we can also derive other predictions in a different scale (the
one of the solar system, for instance).

In order to establish what would be the cosmological effect corresponding
to the gravitational potential of a particle within an expanding universe
we need to know what the averaged cosmological density of matter,
corresponding to some epoch, say $B$, is as a function of the  difference of
the
cosmological time with respect to the epoch of reference, say $A$
(B at time $t_{B}$ has a gravitational action at a distance on any event point
at $t_{A}$.)
The former will, of course, be written in terms of the space distance $r_{B}$
determined
from causal considerations; i.e., gravity only influences points causally
connected.
Both points, $A$ and $B$, satisfying, therefore,  $ds^{2}_{AB}=0$.

To begin with, let us consider the solution of the standard cosmological model
without pressure
\begin{equation}
ds^{2}=c^{2}dt^{2}-x(t)^{2}d\sigma^{2}
\end{equation}
where the equation of conservation of matter is also satisfied
\begin{equation}
\rho(t_{A})x(t_{A})^{3}=\rho(t_{B})x(t_{B})^{3}
\end{equation}

On the other hand, it is always possible to write down the following relation
arising from the value of Hubble parameter corresponding to the  epoch $t$
\begin{equation}
tH(t)=T(\Omega)
\end{equation}
or equivalently, $\frac{d \log [x(t)]}{dt}=\frac{T(\Omega)}{t}$
(where $T(\Omega)$, the cosmic time, is some constant depending only on the
density parameter
$\Omega$). For the open model $\frac{2}{3}<T(\Omega)<1$ (Actual observational
values are in the range $T\sim 0.75 - 0.9$), thus (3) is readily
integrated to yield $x(t)=(\frac{x(t_{A})}{t_{A}^{T}})t^{T}$;
we have assumed $T(\Omega)\approx T$; $T$ being a constant for the epoch $t$
when $t\approx t_{A}$.
Therefore,
\begin{equation}
\rho(t_{A})(t_{A})^{3T}=\rho(t_{B})(t_{B})^{3T}
\end{equation}
\begin{equation}
ds^{2}=c^{2}dt^{2}-t^{2T}(\frac{x(t_{A})}{t_{A}^{T}})^{2}d\sigma^{2}
\end{equation}
Now, only points causally connected have gravitational influence on A,
therefore, for these points (say $B$), $ds_{AB}^2=0$ and, we should
have for the relative comoving coordinates distance between them:
$r_{B}=\frac{ct_{A}}{x(t_{A})(1-T)}\{1-(\frac{t_{B}}{t_{A}})^{1-T}\}$
or, equivalently,
$\frac{t_{B}}{t_{A}}=(1-\frac{r_{B}}{r_{0}(A)})^{\frac{1}{1-T}}$
where we have  defined the "horizon" of $A$ as
$r_{0}(A)\equiv \frac{ct_{A}}{x(t_{A})(1-T)}$. Now, (4) taken into account, we
have
\begin{equation}
\rho(r_{B})=\frac{\rho(t_{A})}{(1-\frac{r_{B}}{r_{0}(A)})^{\frac{3T}{1-T}}}
\end{equation}

The preceding equation relates the observed effective density of matter
for the isotropic universe as a function of the relative distance $r_{B}$
to a reference space-time point at the space-like slice given by $t=t_{A}$.
The direct consequence we can derive from this is the Poisson equation
for the gravitational  cosmological potential that this effective matter
density is able to produce at $A$
\begin{equation}
\frac{1}{r_{B}}\frac{d^{2}[ r_{B}\phi_{A}(r_{B})]}{dr_{B}^{2}}=4\pi G
\rho(r_{B})
\end{equation}

The only meaningful quantity being its present value
($\phi(A)\equiv \lim_{r_{B}\rightarrow 0}\phi_{A}(r_{B})$)
because for $r_{B}>0$ we evaluate quantities lying on the past of $A$.

We can always choose the integration constants  such that this limit
exists. The final result is
\begin{equation}
\phi(A)=-n(T)\{4\pi G\rho(t_{A})(\frac{ct_{A}}{x(t_{A})})^{2}\}
\end{equation}
where we  have introduced a numerical constant,
$n(T)\approx \frac{0.104}{T^{7/2}}$.

Now, for two nearby points,$A$ and $A'$, we must add to their hamiltonian an
additional  gravitational potential, namely,
\begin{equation}
\Delta \phi\equiv  \phi(A)-\phi(A')=-4\pi
G\rho(t_{A})n(T)\{(\frac{ct_{A}}{x(t_{A})})^{2}
-\frac{\rho(t_{A'})}{\rho(t_{A})}(\frac{ct_{A'}}{x(t_{A'})})^{2}\}
\end{equation}
or, taken into account that
$\frac{\rho(t_{A'})}{\rho(t_{A})}=(\frac{t_{A}}{t_{A'}})^{3T}$,
and $x(t_{A'})=x(t_{A})(\frac{t_{A'}}{t_{A}})^{T}$

\begin{equation}
\Delta \phi=4\pi
%% FOLLOWING LINE CANNOT BE BROKEN BEFORE 80 CHAR
G\rho(t_{A})n(T)(\frac{c}{x(t_{A})})^{2}t_{A}^{2}\{(\frac{t_{A}}{t_{A'}})^{5T-2}
-1\}
\end{equation}
but $\frac{t_{A}}{t_{A'}}=\frac{1}{(1-\frac{r}{r_{0}(A)})^{\frac{1}{1-T}}}$;
furthermore, it is easy to obtain, for $r\ll r_{0}(A)$
\begin{equation}
\Delta \phi=4\pi
%% FOLLOWING LINE CANNOT BE BROKEN BEFORE 80 CHAR
G\rho(t_{A})n(T)c_{A}^{2}t_{A}^{2}\{\frac{(5T-2)}{c_{A}t_{A}}r+\frac{(4T-1)(5T-2
)}{2c_{A}^{2}t_{A}^{2}}r^{2}+ ...\}
\end{equation}
where $c_{A}\equiv \frac{c}{x(t_{A})}$ is the speed of light at $t_{A}$.

Now, upon defining:$\rho_{c}\equiv\frac{3H^{2}}{8\pi G}$, $t_{A}H\equiv T$ and
$\Omega\equiv\frac{\rho(t_{A})}{\rho_{c}}$, and dropping the index $A$, we have
\begin{equation}
\Delta \phi=\frac{3}{2}\Omega H c[n(T)T(5T-2)]\cdot r +O[(rH/c)^{2}]
\end{equation}
It is also easy to verify  that, in spite of the fact
that we still do not know the exact value corresponding to the
cosmic time, we would have, for $\frac{2}{3}< T <1$,
as expected from observational considerations,
$\frac{3}{2}n(T)T(5T-2)\approx\frac{3}{5}$;
finally
\begin{equation}
\Delta \phi\approx\frac{3}{5}\Omega H c \cdot r
\end{equation}
Still, another expression for $\Delta \phi$ in terms of the reduced
Hubble constant (recall $H=H_{o}h=3.2408\cdot 10^{-18}h s^{-1}
=100 km\cdot s^{-1}\cdot Mpc^{-1}\cdot h$) is
\begin{equation}
\Delta \phi=\frac{3\Omega h}{5}[H_{o}c]\cdot r
\end{equation}
Hence, we see that there exists a universal acceleration,
$a_{o}\equiv\frac{3\Omega h}{5}[H_{o}c]$;
its value can be estimated since we know several observational constraints on
$\Omega$
\cite{kn:Peebles},$0.01<\Omega<0.3$
\begin{equation}
5.8\cdot 10^{-12}h (m s^{-2})<a_{o}<1.8\cdot 10^{-10}h (m s^{-2})
\end{equation}

For the newtonian gravitational problem of computing relative motions in
comoving coordinates,
we make the approximation of considering every local point as relatively static
with respect to the global expansion of the universe. We see from (14) that
this
is not clearly appropriate when accelerations are those of the order of
$a_{o}$,
and, in any case this additional constant correction should be taken into
account to get the
right motion. Equation (14) is, therefore, important in order to understand
gravitational dynamics when accelerations are of the order of
$10^{-10}h(ms^{-2})$.
It is easy to verify that, for stars in the gravitational field of galaxies,
the order of accelerations is just that of $a_{o}$ and, therefore,
the cosmological correction might play an important role.

It is now evident that this will
amount to correcting the equation of motion by adding a constant
acceleration toward the centre of local gravitational forces:
\begin{equation}
m\frac{v^{2}}{r}=\frac{GMm}{r^{2}}+ma_{o}
\end{equation}
The equation of motion is just frame depending, since
we now use masses $M$ and $m$ as if they were static in a fictitious
space-like slice of the expanding universe (something analogous to
a "Coriolis force problem"). On the other hand, the very
existence of this acceleration will not change the energy of the field;
in fact, it is very easy to demostrate
that for the corresponding postnewtonian metric,
$g_{00}=1-\frac{2GM}{c^{2}r}+2\frac{a_{o}r}{c^{2}}$,  the Weyl tensor is
not depending on the value of $a_{o}$\cite{kn:Plebanski} and, hence, it does
not
change the
meaning of the singularity at $r=0$.

Now, we are interested in the dynamics of a fluid of matter in
a radial  velocity field given by (16). If the density of matter
is decreasing with distance from the centre (as expected for a finite
cumulative
mass of the
galaxy) , then, at a distance $r$ from the centre,
there should be convective density waves coming
from the existence of two-layer interfaces at relative velocity
given from (16),$\Delta v(r)=\frac{1}{2vr}(v^{2}(r)-\frac{2GM}{r})\Delta r$,
where $v\approx \frac{1}{2}(v_{1}+v_{2})$ is the averaged velocity
corresponding
to the two layer's fluid interface at $r$. If $\rho(r+\Delta r)<\rho(r)$ then
the problem is known as the Kevin-Helmholtz instability. For this problem there
will be density perturbation waves. These will
be unstable under the following propagation condition for the wave number $k$
(see, for instance,\cite{kn:Weissman} and references therein),
\begin{equation}
k>\frac{g(\alpha_{2}-\alpha_{1})}{\alpha_{2} \alpha_{1}(\Delta v)^{2}}
\end{equation}
where $g=\frac{GM}{r^{2}}+a_{o}$ is the gravitational acceleration and
$\alpha_{1}=\frac{\rho(r+\Delta r)}{\rho(r)+\rho(r+\Delta r)}$,
$\alpha_{2}=\frac{\rho(r)}{\rho(r)+\rho(r+\Delta r)}$
are the relative densities corresponding to the layers.

We now notice that for $v^{2}(r_{o})\rightarrow\frac{2GM}{r_{o}}$,
($\alpha_{1}<\alpha_{2}$) the minimum value of $k$ tends to infinite,
therefore,
there would exist unstable density waves only
if $\alpha_{1}\rightarrow\alpha_{2}$ or equivalently if,$\rho (r_{o}+\Delta
r)\rightarrow\rho (r_{o})$
and, from the conditions of the problem, this density
will get a constant minimal value for that distance. Hence, (16) taken into
account,
the radial distance given by
\begin{equation}
r_{o}=(\frac{GM}{a_{o}})^{\frac{1}{2}}
\end{equation}
should determine the surface of minimal density for
the galaxy on which unstable density wave perturbations are allowed.

Let us write the equation of motion (16) as if it would exist a
fictitious "dark matter spherical halo"
\begin{equation}
v^{2}=\frac{G}{r}\{{M+M_{DM}}\}
\end{equation}
where "dark matter halo mass" is given by $M_{DM}(r)=\frac{a_{o}r^{2}}{G}$.
It is a remarkable fact that for $r_{o}$ the "fictitious spherical halo" mass
satisfies
$M_{DM}(r_{o})= M$, which seems to be confirmed by observations of ratios of
"dark matter halos" masses versus luminous masses in galaxies when the optical
radius, $r_{25}$ is reached\cite{kn:Bahcall}.
Thus,very approximately we should have for galaxies
\begin{equation}
r_{o}\approx r_{25}
\end{equation}
which is also consistent with the minimal density character of $r_{o}$.

The key point is now the existence of a limit for circular motion velocities:
\begin{equation}
v_{\infty}\approx v(r_{o})\approx (4GMa_{o})^{1/4}
\end{equation}
which allows us to write the following equation for the mass of a galaxy
in terms of its limiting velocity:

\begin{equation}
M=\frac{v_{\infty}^{4}}{4 a_{o} G}
\end{equation}

The preceding equation was already deduced from the empirical theory of gravity
of M. Milgrom and J. Bekenstein \cite{kn:Milgrom1}; in such a theory, these
authors
propose an empirical modification for Newton's law of gravity; the main
hypothesis concerns also the value of accelerations with respect to a "new"
universal constant, $a_{o}^{M-B}$.
In spite of the fact that they did not obtain
its value  from first principles, they were able to get a tentative value for
it
from the observed masses to limiting velocity ratios of spiral galaxies.
They also suggested the existence of an eventual
"machian-like" relation between the value of such a constant and that
corresponding to the product of the speed of light
and Hubble constant $a_{o}^{M-B}\approx h\cdot H_{o}\cdot c$. In this paper we
have produced a confirmation of this suggestion (notice
$a_{o}^{M-B}=4\cdot a_{o}\sim h \cdot H_{o}\cdot c$).
Nonetheless, here, we have not made use of drastic modifications of gravity,
yet we have obtained the essential features that would explain the nature of
rotation
curves in galaxies just from (several solutions of) general relativity,
thus, pretty confirming again this theory of gravitation.

We could obtain a tentative value of the constant $a_{o}$ from our local
estimates
corresponding to rotation velocities of bodies in our galaxy( see
\cite{kn:Rubin} and references therein).

{}From the  cosmological correction to local newtonian motion we have,
$v^{2}_{TH}=\frac{GM(r)}{r}+a_{o}r$. Now, we can try a $\chi^{2}$ fit between
this and experimental values. In
order to do it we have to consider first a distribution of matter arround the
centre
which could be modelled by the use of a variable density of mass of the form:
$\rho(\vec{r})\approx\rho_{o}e^{-R/b}sech^{2}(\frac{z}{z_{0}})$
($\vec{r}$ in cilindrical coordinates), hence,
$M(\vec{r};\rho_{o},b,z_{0})=\int_{0}^{\vec{r}}\rho(\vec{r})d\omega(\vec{r})$
($\rho_{o}$, $b$ and $z_{0}$ being constants.) From this, we look for the
minimum value of
%% FOLLOWING LINE CANNOT BE BROKEN BEFORE 80 CHAR
$\chi^{2}(\rho_{o},b,z_{0};a_{o})=\sum\frac{(v_{TH}(r_{i})-v_{exp}(r_{i}))^{2}}{
\sigma^{2}_{i}}$,
where the values are weighted by their correspondig experimental uncertainty
($\sigma \approx 50 km/s$.)

The $\chi^{2}$ minimum (see Figure 1) has been achieved for the following set
of
parameters:
$\rho_{o}\approx 0.31 M_{\odot} ps^{-3}=2.1 \cdot 10^{-20} kg \cdot m^{-3}$
($M_{\odot}$ a solar mass.),
$b\approx 7.0 kps$, $z_{0}\approx 1.5 kps$ and $a_{o}\approx (2.0 \pm 0.6)
\cdot
10^{-11} ms^{-2}$.
This procedure leads, of course, to a value  of $a_{o}$ which is
in the range of those already predicted from previous estimates of $\Omega$.
Thus,  we have a new and independent estimate for the value
of $\Omega h$
\begin{equation}
\Omega h=0.036 \pm 0.01
\end{equation}

If matter were only baryonic, i.e. $\Omega\leq 0.2$, then, also  very likely,
we
could
determine independently a value corresponding to this parameter
from the known elements abundances, i.e., $\Omega$ should be given
from primordial nucleosynthesis calculations\cite{kn:Walker}.
\begin{equation}
\Omega h^{2}=0.0125\pm 0.0025
\end{equation}
And, since we have got a different value for $\Omega h$, it
is also possible to determine  $h\approx0.35\pm 0.16$;
this seems to favour a low value for Hubble's constant in
accordance with some recent calculations \cite{kn:Sandage}.

It is also very easy to make use of empirical data  obtained by Milgrom
for values of masses given by equation  (22)  in order to compute an
alternative value for $a_{o}$;
$a_{o}^{M-B}=4a_{o}\approx 8\cdot 10^{-10}h^{2}(ms^{-2})$\cite{kn:Milgrom2}.
Therefore,$a_{o}\approx 2\cdot 10^{-10} (ms^{-2})  h^{2}\approx 2\cdot 10^{-11}
(ms^{-2})$
which also implies $h\approx 0.31$, again a low value for Hubble's constant.

On the other hand, we have got for our Galaxy, the following numbers:
$M\equiv\lim_{r\rightarrow \infty} M(r)\approx 3.4\cdot 10^{11}M_{\odot}$
and $r_{o}\approx 49 kps$.

We can produce the following  relation for
the value of the mass of a galaxy and the corresponding limiting velocity
(in $km\cdot s^{-1}$)
\begin{equation}
\frac{M}{M_{\odot}}\approx 93.7\cdot v_{\infty}^{4}
\end{equation}

Now, upon considering a solar luminosity relation
$\frac{M_{\odot}}{L_{\odot}}=1$,
this is equivalent to $-M_{B_{T}^{0}}=-2.5\log[<\frac{M}{L}>]
+22.21+10\cdot(\log[v_{\infty}]-2.2)$;
which is the largely used empirical relation of Tully-Fisher (see
\cite{kn:Fukigita}
and \cite{kn:Tully}.)
We now use this theoreticaly derived equation in order to test the observed
limiting velocity  ratio corresponding to different types of spiral galaxies of
the same luminosity,
(Sa=$<\frac{M}{L}>\approx 3.1$ and Sc=$<\frac{M}{L}>\approx 1.2$, for
instance);
the value obtained from the previous expression is given approximately by
$v_{\infty}^{Sa}=1.3v_{\infty}^{Sc}$ which explains quite accurately the
observed ratio.

Similarly, it is also possible to write, for the diameter of
a galaxy ($D_{o}\approx 2\cdot r_{o}$) expressed in kpc,
\begin{equation}
\frac{M}{M_{\odot}}=3.5\cdot 10^{7}D_{o}^{2}
\end{equation}
And, therefore, we finally get, also for astrometric purposes,
$-M_{B_{T}^{0}}=14.15 +5\log(D_{o})-2.5\log[<\frac{M}{L}>]$
which should be compared with the empirical approximate relation used for
spiral galaxies\cite{kn:Girardi}, $-M_{B_{T}^{0}}=12.4 +5.7\log(D_{25})$
and the one obtained for elliptical galaxies\cite{kn:Giuricin},
$-M_{B_{T}^{0}}=13.8 +4.8\log(D_{25})$;
$D_{25}$ being the visible diameter. It is a remarkable fact that both
expressions seem to be very closely related with the theoretical one we have
obtained from (26).

In summary, we have given a first principle's solution to the problem
of understanding rotation curves of galaxies without the introduction
of (undetected) dark matter. The solution implies that there should
exist a constant cosmological acceleration implying a relative
delay for the large scale gravitational interaction  computed
for nearby points. This acceleration should be added in order to correct
the equation of motion when a constant time comoving coordinate frame is used
for
the system.
Moreover, there  would exist  absolute effective limits
for galaxies where unstable gravity waves are allowed,this is the minimal
density surface. We gave it as an expression in terms of cosmological
parameters; therefore, we should, hereafter, consider the
galaxies not only as astrophysical individual objects but also as having
several
new unexpected cosmological properties. We have seen that, by considering
our local galaxy motions and this cosmological acceleration
in Newton's law of gravity, it is possible to obtain that the value of the
closure parameter is  less than $1$.
If we considered  primordial nucleosynthesis estimates, this conclusion is also
consistent with a low value for  Hubble's constant, which seems to be
in accordance, after Milgrom's empirical relation, with the observed
non-decreasing of rotation velocities in galaxies.
The well known Tully-Fisher relation has been obtained as a consequence of the
"machian" character of the limiting velocity for a gravitational field;
this seems to be, therefore,
the first time Mach's principle has empirical observable consequences.
Several new tests of the existence of the cosmological
correction to local newtonian equation of motion in our solar system are in
preparation.

The author whishes to thank  Drs. Jose Luis Sanchez-Gomez,  Rosa Dominguez
(Universidad auton"noma de Madrid, Spain)
and Joaquin Diaz-Alonso (Observatoire de Meudom, France), for stimulating
discussions.
The search for a Machian-like gravitational
potential was envisaged, in the framework of quantum mechanics, by  Joaqu!n
Diaz-Alonso.

\begin{figure}
\begin{center}
\leavevmode
\epsfxsize=290.truept
\epsffile{curve.eps}
\end{center}

\caption{$\chi^{2}$ fit to the Galaxy rotation curve data.}
\label{eqzz}
\end{figure}

\end{document}